\documentclass[fleqn,10pt]{wlscirep}
\usepackage[utf8]{inputenc}
\usepackage[T1]{fontenc}
\usepackage{lineno}

\title{Deciphering Bitcoin Blockchain Data by Cohort Analysis}

\author[1,4,+]{Yulin Liu}
\author[2,1,+,*]{Luyao Zhang}
\author[3,1,+]{Yinhong Zhao}

\affil[1]{SciEcon CIC, London, WC2H 9JQ, United Kingdom}
\affil[2]{Data Science Research Center and Social Science Division, Duke Kunshan University, Kunshan, Jiangsu, 215316, China}
\affil[3]{Duke University, Durham, NC, 27708, United States}
\affil[4]{Bochsler Finance, Zug, 6300, Switzerland}

\affil[*]{Corresponding author: email: lz183@duke.edu, address: Duke Kunshan University, No. 8 Duke Avenue, Kunshan, Jiangsu, 215316, China}

\affil[+]{The authors are listed in alphabetical order according to last names and contributed equally to this work.}

\begin{abstract}
Bitcoin is a peer-to-peer electronic payment system that has rapidly grown in popularity in recent years. Usually, the complete history of Bitcoin blockchain data must be queried to acquire variables with economic meaning. This task has recently become increasingly difficult, as there are over 1.6 billion historical transactions on the Bitcoin blockchain. It is thus important to query Bitcoin transaction data in a way that is more efficient and provides economic insights. We apply cohort analysis that interprets Bitcoin blockchain data using methods developed for population data in the social sciences. Specifically, we query and process the Bitcoin transaction input and output data within each daily cohort. This enables us to create datasets and visualizations for some key Bitcoin transaction indicators, including the daily lifespan distributions of spent transaction output (STXO) and the daily age distributions of the cumulative unspent transaction output (UTXO). We provide a computationally feasible approach for characterizing Bitcoin transactions that paves the way for future economic studies of Bitcoin. 
\end{abstract}
\begin{document}

\flushbottom
\maketitle

\thispagestyle{empty}

\section*{Background \& Summary}

Bitcoin is a peer-to-peer electronic payment system that has rapidly grown in popularity in recent years~\cite{nakamoto_2008_bitcoin, bhme_2015_bitcoin, halaburda_2018_blockchain, subacchi_2021_from}. As a distributed ledger technology (DLT), Bitcoin records newly generated transactions in a decentralized way, eliminating the need for intermediaries like banks and reducing transaction costs~\cite{ ornes_2019_core, townsend_2020_distributed, harvey_2021_defi}.

\begin{figure}[ht]
    \centering
    \includegraphics[width = \linewidth]{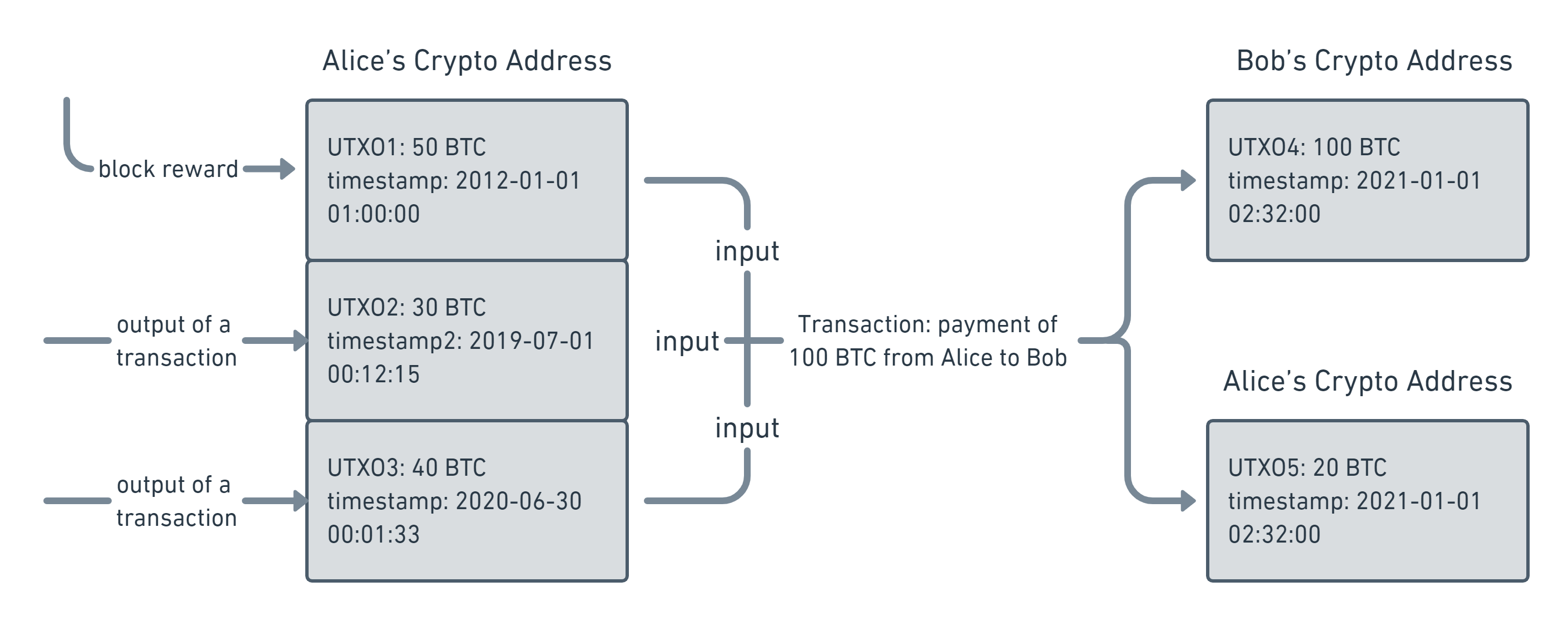}
    \caption{An example of UTXO birth and death. UTXOs 1, 2, and 3 were spent in a transaction taking place between Alice and Bob and were transformed to UTXOs 4 and 5. UTXOs 1, 2, and 3 became STXOs after the transaction.}
    \label{fig:example}
\end{figure}

Bitcoin relies on recording the unspent transaction outputs (UTXO) to efficiently verify newly generated transactions~\cite{delgadosegura_2019_analysis, zahnentferner_2018_chimeric, urquhart_2016_the, chakravarty_2020_the, prezsol_2019_another}. An illustrative example of UTXO is shown in Figure~\ref{fig:example}. A UTXO can be generated either as block rewards or outputs of transactions. Block rewards are newly minted bitcoins (BTC) distributed to miners for their work to maintain the network, such as routing transactions and validating blocks. In fact, all UTXOs can be dated back to block rewards. The timestamp is recorded when a UTXO is generated. A UTXO is spent and converted into a spent transaction output (STXO) when it is used as the input of a transaction. A timestamp is again recorded when the UTXO is spent, and each UTXO can be spent only once. Such a unique feature allows us to calculate the age of each UTXO and the lifespan of each STXO as we do in population data. Take Figure~\ref{fig:example} as an example. As of July 1, 2020, UTXOs 1-3 are 8.5-years, 1-year, and 1-day old, respectively. Immediately after Alice’s payment to Bob on January 1, 2021, UTXOs 1-3 are converted to STXOs with ages of  9 years, 1.5 years, and 0.5 years and 1-day old, respectively.

\par
Noticing the unique structure of the Bitcoin blockchain data, we apply cohort analysis~\cite{glenn_2005_cohort, mason_1973_some, breslow_1983_multiplicative, jiang_2016_cohort, omidvartehrani_2018_cohort}, originally developed for population data, to analyze it. To continue the analogy with the population data, we say a UTXO is born when it is generated as block rewards or the output of a transaction, and we say a UTXO is dead when it is spent as the input of another transaction. In this way, all UTXOs generated on the same day form a daily birth cohort, and all UTXOs spent on the same day form a daily death cohort. We define the age of a UTXO as the difference between “now” (the date on which we are working) and the time when it was born. We define the lifespan of an STXO as the difference between the time when the STXO was dead and the time when it was born. Thus, all UTXOs within an age range form an age cohort, and all STXOs within a lifespan range form a lifespan cohort. With this framework, we naturally replicate in Bitcoin blockchain data a trinity of birth, death, and age cohorts using population cohort analysis.
\par
Usually, we need to query the complete history of Bitcoin blockchain data to acquire variables with economic meaning. With over 1.6 billion historical transactions on the Bitcoin blockchain, it has become increasingly difficult and computationally intensive now to download the complete Bitcoin blockchain records. It is thus important to query Bitcoin transaction data in a way that is more efficient and provides economic insights~\cite{liu_2021_cryptocurrency}. Cohort analysis provides a new perspective from which we can analyze data within each cohort separately before integrating them into a time series.
\par
\begin{figure}[ht]
\centering
\includegraphics[width=\linewidth]{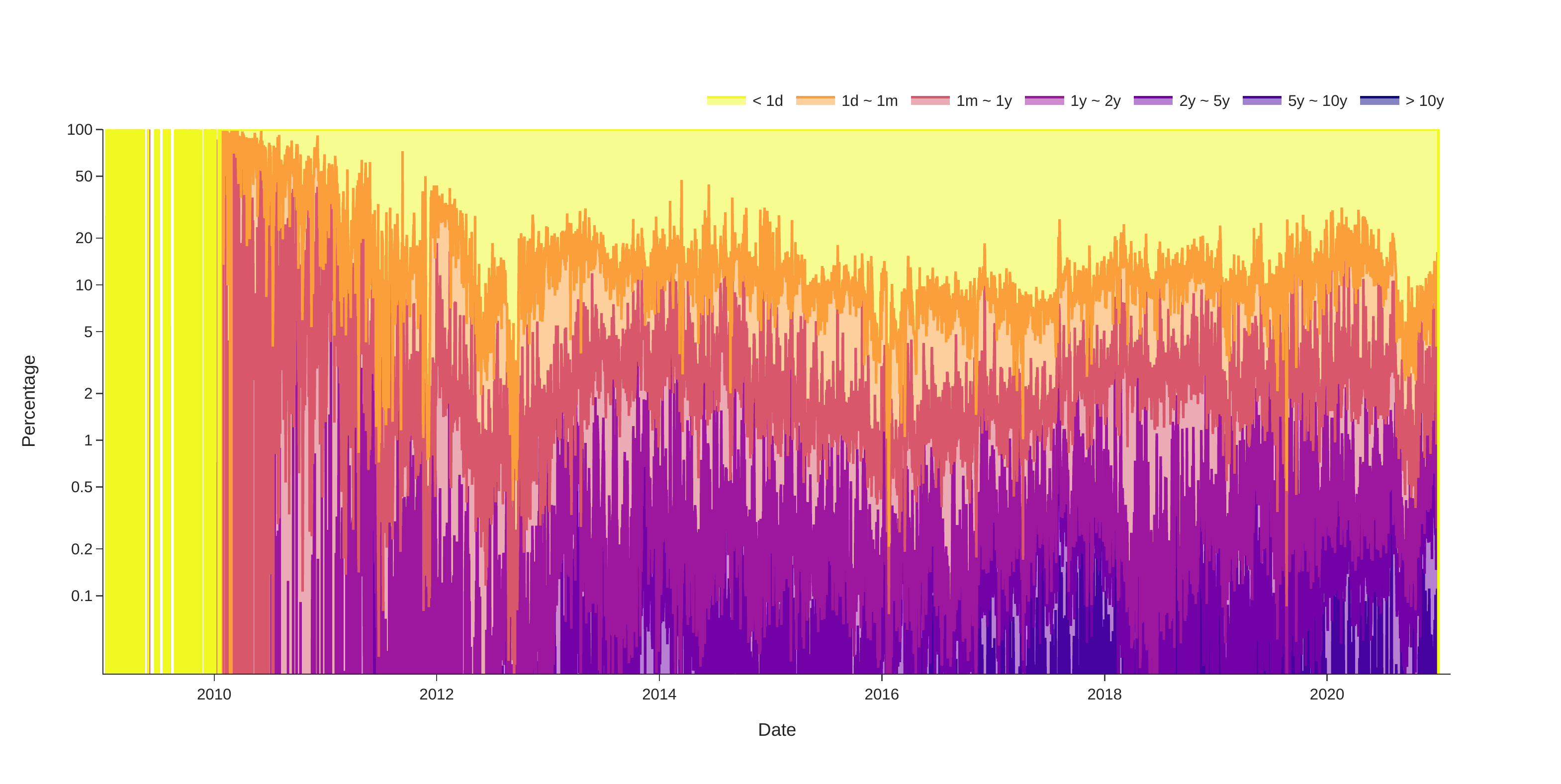}
\caption{Lifespan distribution of BTC STXOs. The figure shows the log percentage of spent transaction outputs with different lifespans in each day until Feb. 2021. For example, by Feb. 2021, the STXOs with lifespans of less than one day accounted for 80\% of all STXOs, while those with lifespans between 1 day and 1 month accounted for another 15\%.}
\label{fig:1}
\end{figure}
\begin{figure}[ht]
\centering
\includegraphics[width=\linewidth]{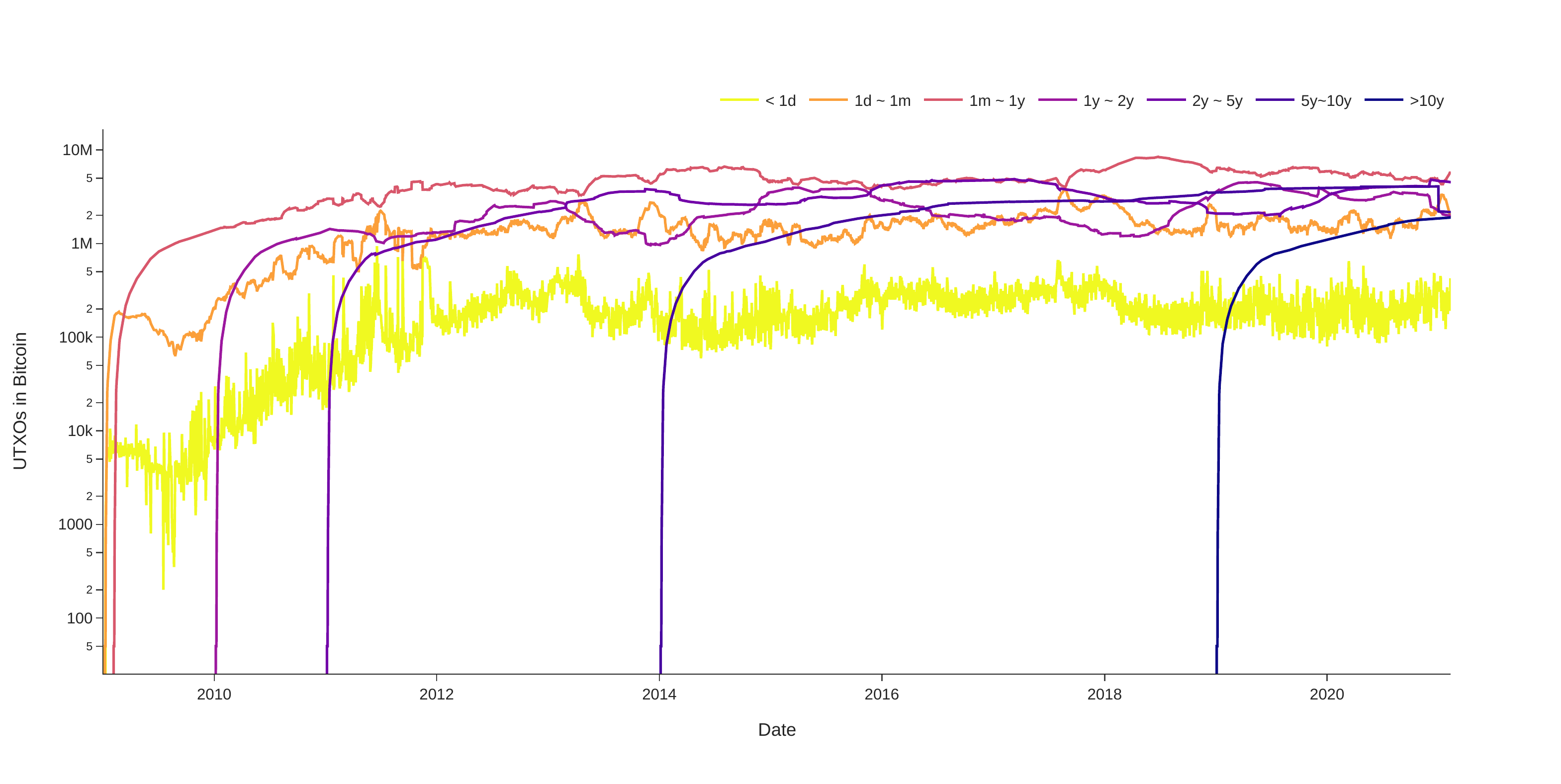}
\caption{Number of BTC UTXOs by age. The figure shows the total cumulative unspent transaction outputs by age. For example, by Feb. 2021, there were approximately 200k UTXOs less than 1 day old used as the medium of exchange and approximately 2 million UTXOs more than 10 years old lost or used as store of value.}
\label{fig:2}
\end{figure}

Our workflow is displayed in Figure~\ref{fig:cohort}. We query and process Bitcoin transaction input and output data within each daily cohort. By doing so, we successfully create datasets and visualizations for some key Bitcoin transactions indicators, including the daily lifespan distributions of STXOs as percentages (Figure \ref{fig:1}) and the cumulative daily age distributions of UTXOs (Figure \ref{fig:2}). These visualizations can be used to study the functions of bitcoin (BTC) as a currency. The three functions of a currency include acting as a store of value, unit of account, and medium of exchange. For example, Figure \ref{fig:2} shows the number of BTCs in UTXOs (i.e., BTCs that have not been spent) by age distribution. By the end of 2020, approximately 2 million BTCs had not been transacted for more than 10 years. In the past 5-10 years, 2-5 years, and 1-2 years, approximately 2 million, 4.5 million, and 3 million BTCs, respectively, remained inactive. This equals approximately 11.5 million BTCs not having been transacted for more than 1 year. These BTCs serve as a time deposit and act as a store of value. Moreover, approximately 5 million BTCs are alive for 1 month to 1 year. These BTCs are similar to a demand deposit. Frequently transacted BTCs are those with ages between 1 day and 1 month (2 million) and less than 1 day (0.2 million). These BTCs act as a medium of exchange.
\par
Our final datasets include one dataset that characterizes STXOs and one that characterizes UTXOs, which are both smaller than 1 MB. Moreover, cohort analysis keeps data querying and processing to a minimum for future updates and enables automated updates. We thus provide a computationally feasible approach for characterizing BTC transactions, which paves the way for future economic studies of Bitcoin. Our methods can be generally applied to other cryptocurrencies that adopt UTXO protocols, including Litecoin, Dash, Zcash, Dogecoin, and Bitcoin Cash.

\section*{Methods}

While the Bitcoin transaction output data are publicly available on its blockchain, we find the size of the raw data (approximately $1.3$ TB) overwhelming to process, even with cloud computing platforms. To improve the efficiency of computation, we first retrieve the data relevant to the study to create a more manageable data table of only $45$ GB. By partitioning this data table into daily birth and death cohorts, we can analyze the STXOs and UTXOs in each cohort separately to summarize the daily characteristics of transaction outputs and create visualizations based on the cohort summary. Our method can be adapted to the creation of future blocks — we only need to process the transaction output data from the latest cohort and append the summary to the current version.
\par
\subsection*{Creating Partitioned Tables}
\par
Our primary workplace is Google Colaboratory (Colab), a Jupyter Notebook hosted environment from Google, and BigQuery, a data warehouse from Google Cloud Platform. We first query the columns of interest from the public dataset \textit{crypto-bitcoin} on BigQuery~\cite{day_2019_introducing}, which includes the input and output data of Bitcoin. We then join the data queried from input and output data to create a data table that includes the value of UTXO (\textit{value}), the timestamp when the UTXO was created (\textit{block\_timestamp}), and the timestamp when the UTXO was spent as an input of another transaction (\textit{spent\_block\_timestamp}) (this column is left null if the transaction output is unspent). As the UTXO in a transaction is counted in \textit{satoshi} ($1$ satoshi $= 10^{-8}$ BTCs), the actual number of BTCs in a UTXO can be computed by $\#_{UTXO} = value*10^{-8}$, where the $value$ represents the number of BTCs in satoshi. We rely on this derived data table ($1.6$ billion rows, $45$ GB) to conduct further analysis.
\par
To save the cost of the query, we create two partitioned tables based on the derived data table, one by the date in \textit{block\_timestamp} and one by the date in \textit{spent\_block\_timestamp}. This means that the data entries are partitioned either by the date when the UTXOs were created or by the date when the UTXOs were spent. In this way, the program queries only the entries with timestamps in a specific range, which saves a notable amount of computational power. This step can significantly improve query performance and reduce query cost \cite{a2021_introduction}.
\par
\subsection*{Querying and Processing Cohort Data}
\par
The data structure of partitioned tables coincides with our need to process cohort data. The table partitioned by date in \textit{block\_timestamp} naturally divides the derived data into birth cohorts that include the segment of transaction outputs created on the same date, and the table partitioned by date in \textit{spent\_block\_timestamp} divides the derived data into death cohorts that include the segment of transaction outputs spent on the same date. 
\par
\begin{figure}[ht]
    \centering
    \includegraphics[width = \linewidth]{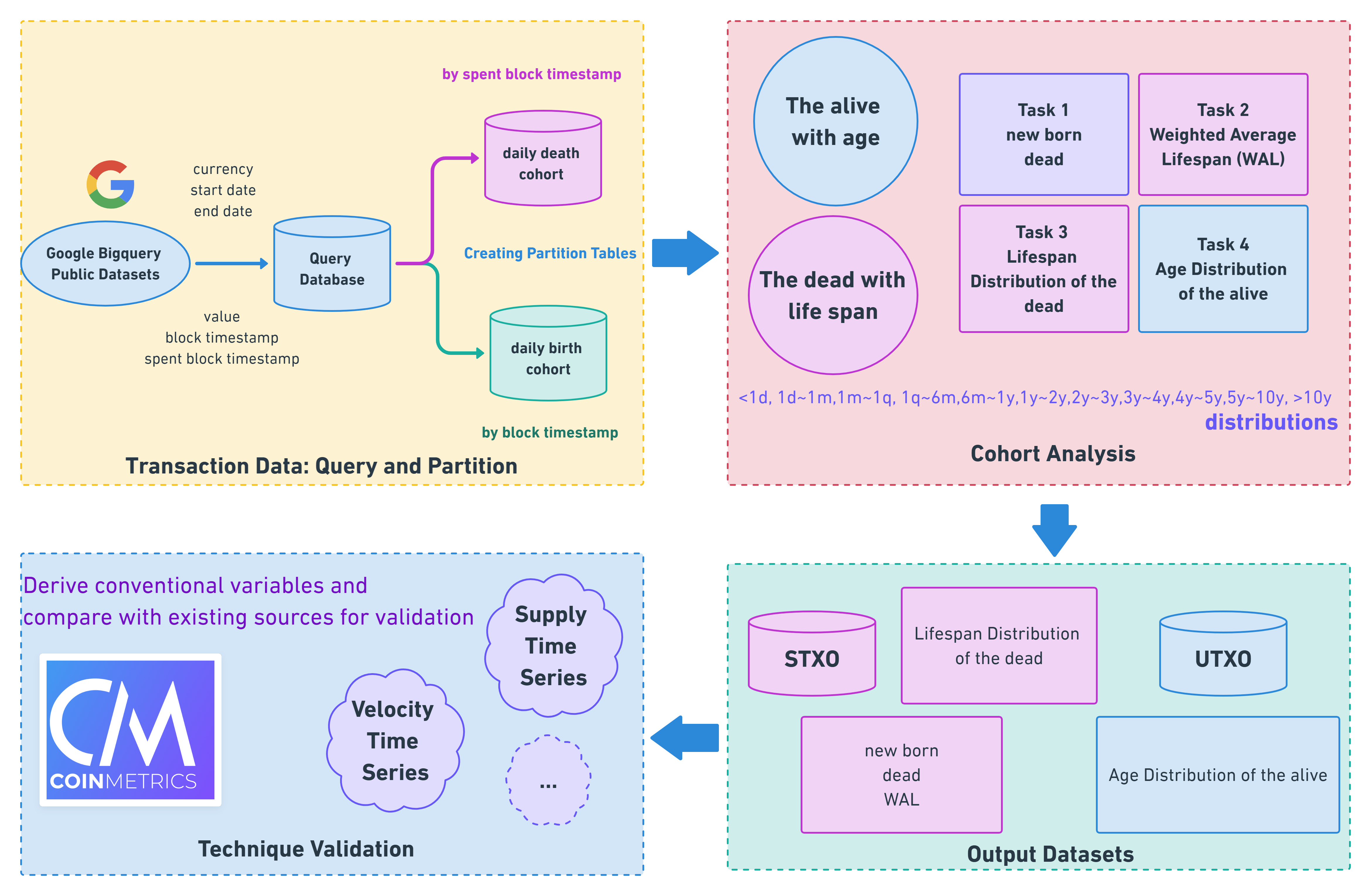}
    \caption{Workflow of cohort analysis on BTC UTXO data}
    \label{fig:cohort}
\end{figure}

We query and process each birth cohort and each death cohort with a loop program following the procedure described in Figure~\ref{fig:cohort}. For each specific date after 2009-01-03, when the first block of Bitcoin was created, the birth cohort data and the death cohort data of that date are queried and imported to Colab from BigQuery. As in Task 1, we compute the total number of BTCs in UTXOs created and spent on that date by summing the number of BTCs in UTXOs in the birth cohort data and the death cohort data respectively. Task 2 focuses on the weighted average lifespan (WAL) on the date, defined as the average lifespan (the difference between the time when the output was spent and the time when the output was created) weighted by the number of BTCs contained in the transaction outputs. WAL can be computed from the death cohort data by the formula: $$WAL[date = i] = \sum_{date=i} (\#_{UTXO} \times Lifespan)/\sum_{date=i} \#_{UTXO},$$ where $Lifespan = spent\_block\_timestamp - block\_timestamp$. 
\par
As in Task 3, we compute the distribution of lifespan with death cohort data on that date by first categorizing UTXOs based on lifespan and then summing the number of BTCs in UTXOs in each category. In Task 4, we apply a more complicated partitioning method to compute the age distribution for each specific date. The age of a UTXO is defined as $age = working\_date - block\_timestamp$, where working date means the date of interest for the data cohort being studied. Each UTXO that remains alive on a specific date must satisfy both conditions: a) its \textit{block\_timestamp} must be smaller than the end of the working date, which means that the UTXO was created sometime before or on the date, and b) its \textit{spent\_block\_timestamp} must either be null, which means the UTXO was not spent before 2021-02-10, or be larger than the end of the working date, which means that the UTXO was spent sometime after the working date but before 2021-02-10. Thus, we cannot simply interpret this information as either birth or death cohort data. Instead, we must first query the data needed to compute the age distribution for a twelve-month or six-month period depending on the size of the data in each year and then split the queried data into daily cohorts in the Python program. We compute the age distribution of each daily cohort by categorizing the age of each UTXO and summing up the number of BTCs in UTXOs in each category.
\par

\subsection*{Visualizing the Time Series}

The result of our analysis is condensed into time-series data that include the number of BTCs in UTXOs created and spent, the weighted average lifespan, the lifespan distribution, and the age distribution on each date from 2009-01-03 to 2021-02-10. Many visualizations can potentially be generated from this informative time series. For example, BTC token velocity, which we define below as the number of BTCs spent in the last 30 days divided by the circulating supply of BTCs, can be computed by $$v[date = i] = \frac{\sum_{j=0}^{29}\#_{spent}[date=i-j]}{Supply[date=i]}. $$
\par
Our method can be adapted to the creation of future blocks. The time-series data for the past dates are not subject to changes as new blocks are created. As time goes on, we need only query and process the latest data cohorts to extend the time series. We will update the visualizations according to the latest development of Bitcoin, and researchers may easily repeat our work in part or in whole based on their needs.

\section*{Data Records}

The final data records are stored and published on the Harvard Dataverse~\cite{liu_2021_replication}. The records consist of the UTXO and the STXO datasets in csv format. The metadata information of the two datasets is presented in Appendix Tables~\ref{table:A1} and \ref{table:A2}. Data ranges from 2009-01-03 to 2021-02-10, and the data frequency is daily ($n=4421$). The timezone used in the data is UTC+0. In addition to examining Bitcoin, we apply the same cohort analysis to five other cryptocurrencies and generate twelve datasets in total. Detailed information on these data files is presented in Appendix Table~\ref{tab:datafiles}.

\section*{Technical Validation}

\begin{figure}[ht]
\centering
\includegraphics[width=\linewidth]{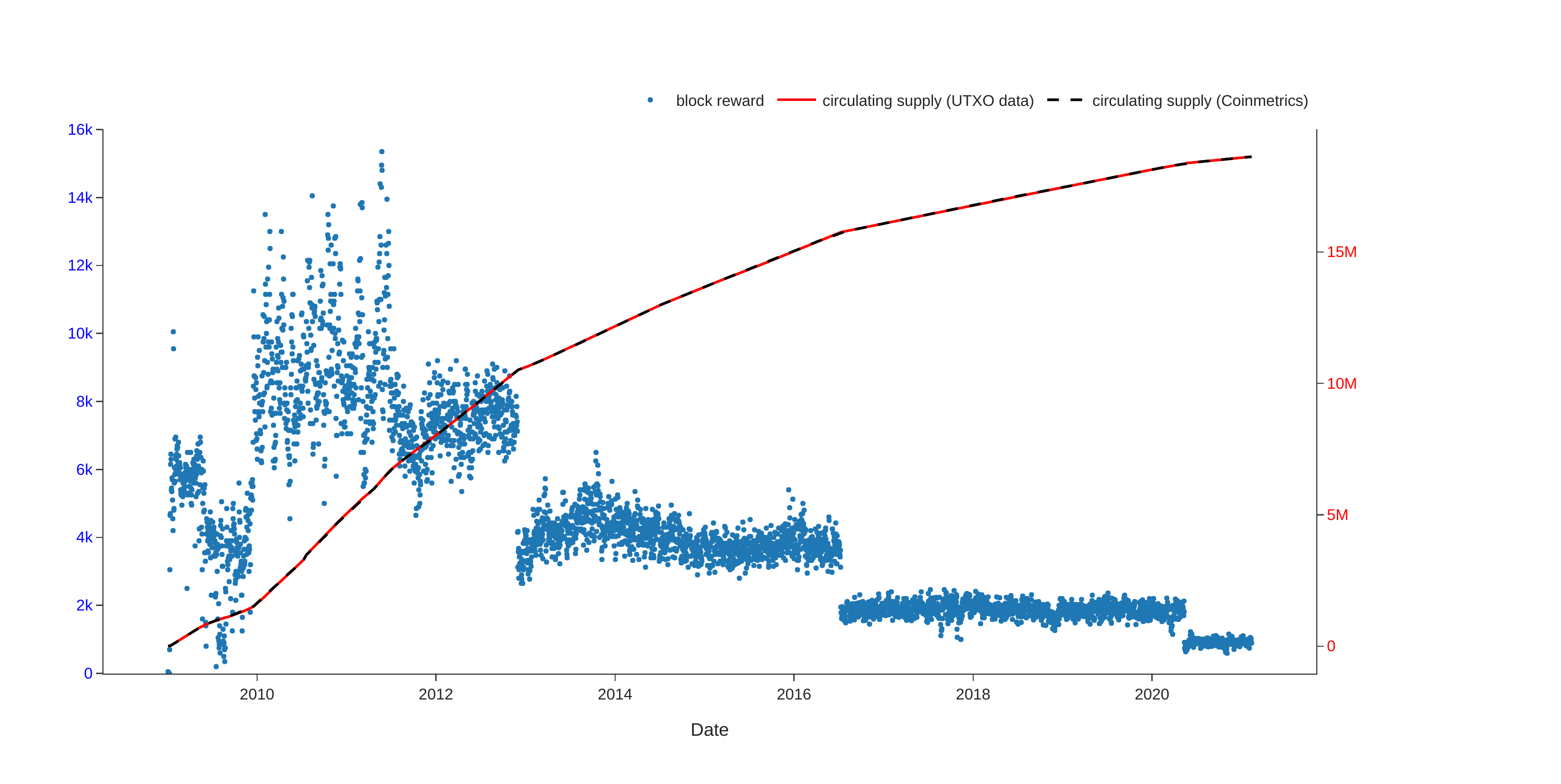}
\caption{Block reward and circulating BTC supply. The figure shows that the block reward (net BTC supply) roughly corresponds to the halving pattern written in the Bitcoin white paper. It also shows that the circulating supply we compute with our UTXO data coincides exactly with the circulating supply from Coinmetrics.}
\label{fig:3}

\end{figure}

To further verify the validity of our methods, we use our data to calculate other variables, including block reward and circulating supply of BTCs, and check whether the results are consistent with descriptions in the Bitcoin white paper~\cite{nakamoto_2008_bitcoin} and external data sources. We compute the circulating supply of BTC by computing the cumulative net new UTXOs with the formula $$Supply[date = i]=Net\_new[date \leq i] = \sum_{date\leq i} \#_{created} - \sum_{date\leq i} \#_{spent}.$$ 
Figure \ref{fig:3} visualizes the block rewards and the circulating supply. Block rewards are the BTC awarded to the miner who wins the right to record a block of transactions by proof-of-work. Supply of the BTCs originates from the block rewards, so the cumulative sum of block rewards is the total number of BTCs in UTXOs, i.e., the circulating supply of BTC. The Bitcoin block reward was initially set at 50 BTCs per block in 2009, which means approximately 7,200 newly minted BTCs every 24 hours. The block reward halves every 210,000 blocks, roughly every four years, until the total BTC supply reaches 21 million~\cite{nakamoto_2008_bitcoin}. As of the time of writing, the daily block reward amounts to approximately 900, and the circulating BTC supply is 18.9 million.

In addition, we calculate the circulating supply of BTCs by summing all UTXOs in different age cohorts because existing BTC are essentially just UTXOs of different ages. We then compare the circulating supply we compute with the circulating supply data obtained from CoinMetrics, a widely used blockchain database~\cite{coinmetrics_2021_coinmetricsdata}. As shown in Figure \ref{fig:3}, the two measures of circulating supply match exactly with each other. Hence, the validity of our data is verified.

\section*{Usage Notes}


\subsection*{Applicability}
Our data can inspire research in finance, computer science, and macroeconomics. Our data can produce new technical indicators for financial studies to predict cryptocurrency bubbles~\cite{shu_2020_realtime, li_2019_sentimentbased}, measure cryptocurrency volatility and systematic risk~\cite{giudici_2019_vector, giudici_2021_libra}, design investment strategies~\cite{pagnottoni_2019_neural, karalevicius_2018_using, resta_2020_technical} and implement portfolio managements~\cite{daniel_2020_the, griffin_2020_is}. For instance, Liu and Zhang~\cite{liu_2021_cryptocurrency} used our data to design automated trading strategies for BTC investment that outperform conventional approaches. In computer science, we can apply the UTXO and STXO data to evaluate blockchain security and scalability~\cite{croman_2016_on, gervais_2016_on, pagnotta_2021_decentralizing}. Wang et al.~\cite{wang_2021_ess} cite our data to demonstrate the scalability issues of the BTC blockchain and propose an efficient storage scheme. Our data can also contribute to event studies that evaluate the effect of macro policies on BTC transactions~\cite{karau_2021_monetary}.  

\subsection*{Limitations and future research}
In this section, we identify the limitations of our current results and directions for future research. 
First, although the frequency of our data is on a daily level, our cohort analysis can produce data with higher frequencies. Table~\ref{tab:blocktime} shows several other cryptocurrencies to which our methodology can be easily applied. The granularity of the data can reach different levels (75 seconds to 10 minutes) depending on the block time of each cryptocurrency.

Second, the age distribution of UTXOs is a limited measure for BTC as a store of value. UTXOs might accumulate ages for at least two reasons other than being a store of value: First, the owner of the UTXOs has lost the private key, or second, the amount of UTXOs in the owner’s account is less than the transaction fee. Owners do not transact these dust UTXOs for cost-benefit reasons. In neither case is age accumulation a sign that BTC acts as a store of value. However, scientific methods to identify the two types of UTXOs have yet to be found. 

Third, the cohort analysis we designed and implemented was for UTXO-based blockchains. However, account-based blockchains, such as Ethereum, Polkadot, and Dfinity, adopt a different accounting method. In the UTXO model, crypto tokens are akin to banknotes issued by central banks; in the account model, crypto tokens are akin to balances in commercial bank accounts. Future research could extend the cohort analysis for account-based blockchains. Moreover, Ethereum, a Turing-complete blockchain, has two types of accounts: externally owned accounts (EOA) and contract accounts, which can be analogized to private and corporate accounts in commercial banks. A comparative study of the two accounts by cohort analysis could be an exciting direction for future research.

\section*{Code availability}

The code used for the cohort analysis is available on GitHub~\cite{sciecon_2021_scieconutxo}. The GitHub repository is also archived by Zenodo~\cite{luyaozhang_2021_scieconutxo}, with the code available in Python and written in Google Colab Notebook with Markdown. \\  first release created on Github: 22 Apr 2021;\\
license: GPL-3.0 License

\bibliography{sample}

\begin{thebibliography}{10}
\urlstyle{rm}
\expandafter\ifx\csname url\endcsname\relax
  \def\url#1{\texttt{#1}}\fi
\expandafter\ifx\csname urlprefix\endcsname\relax\def\urlprefix{URL }\fi
\expandafter\ifx\csname doiprefix\endcsname\relax\def\doiprefix{DOI: }\fi
\providecommand{\bibinfo}[2]{#2}
\providecommand{\eprint}[2][]{\url{#2}}

\bibitem{nakamoto_2008_bitcoin}
\bibinfo{author}{Nakamoto, S.}
\newblock \bibinfo{title}{Bitcoin: A peer-to-peer electronic cash system}
  (\bibinfo{year}{2008}).

\bibitem{bhme_2015_bitcoin}
\bibinfo{author}{Böhme, R.}, \bibinfo{author}{Christin, N.},
  \bibinfo{author}{Edelman, B.} \& \bibinfo{author}{Moore, T.}
\newblock \bibinfo{journal}{\bibinfo{title}{Bitcoin: Economics, technology, and
  governance}}.
\newblock {\emph{\JournalTitle{Journal of Economic Perspectives}}}
  \textbf{\bibinfo{volume}{29}}, \bibinfo{pages}{213--238},
  \url{https://doi.org/10.1257/jep.29.2.213} (\bibinfo{year}{2015}).

\bibitem{halaburda_2018_blockchain}
\bibinfo{author}{Halaburda, H.}
\newblock \bibinfo{journal}{\bibinfo{title}{Blockchain revolution without the
  blockchain?}}
\newblock {\emph{\JournalTitle{Communications of the ACM}}}
  \textbf{\bibinfo{volume}{61}}, \bibinfo{pages}{27--29},
  \url{https://doi.org/10.1145/3225619} (\bibinfo{year}{2018}).

\bibitem{subacchi_2021_from}
\bibinfo{author}{Subacchi, P.}
\newblock \bibinfo{journal}{\bibinfo{title}{From gold to bitcoin and beyond}}.
\newblock {\emph{\JournalTitle{Nature}}} \textbf{\bibinfo{volume}{597}},
  \bibinfo{pages}{626--627}, \url{https://doi.org/10.1038/d41586-021-02615-2}
  (\bibinfo{year}{2021}).

\bibitem{ornes_2019_core}
\bibinfo{author}{Ornes, S.}
\newblock \bibinfo{journal}{\bibinfo{title}{Core concept: Blockchain offers
  applications well beyond bitcoin but faces its own limitations}}.
\newblock {\emph{\JournalTitle{Proceedings of the National Academy of
  Sciences}}} \textbf{\bibinfo{volume}{116}}, \bibinfo{pages}{20800--20803},
  \url{https://doi.org/10.1073/pnas.1914849116} (\bibinfo{year}{2019}).

\bibitem{townsend_2020_distributed}
\bibinfo{author}{Townsend, R.~M.}
\newblock \emph{\bibinfo{title}{Distributed Ledgers: Design and Regulation of
  Financial Infrastructure and Payment Systems}} (\bibinfo{publisher}{The MIT
  Press}, \bibinfo{year}{2020}).

\bibitem{harvey_2021_defi}
\bibinfo{author}{Harvey, C.~R.}, \bibinfo{author}{Ramachandran, A.} \&
  \bibinfo{author}{Santoro, J.}
\newblock \emph{\bibinfo{title}{Defi And The Future Of Finance.}}
  (\bibinfo{publisher}{John Wiley}, \bibinfo{year}{2021}).

\bibitem{delgadosegura_2019_analysis}
\bibinfo{author}{Delgado-Segura, S.}, \bibinfo{author}{Pérez-Solà, C.},
  \bibinfo{author}{Navarro-Arribas, G.} \&
  \bibinfo{author}{Herrera-Joancomartí, J.}
\newblock \bibinfo{journal}{\bibinfo{title}{Analysis of the bitcoin utxo set}}.
\newblock {\emph{\JournalTitle{Financial Cryptography and Data Security}}}
  \bibinfo{pages}{78--91}, \url{https://doi.org/10.1007/978-3-662-58820-8_6}
  (\bibinfo{year}{2019}).

\bibitem{zahnentferner_2018_chimeric}
\bibinfo{author}{Zahnentferner, J.}
\newblock \bibinfo{title}{Chimeric ledgers: Translating and unifying utxo-based
  and account-based cryptocurrencies} (\bibinfo{year}{2018}).

\bibitem{urquhart_2016_the}
\bibinfo{author}{Urquhart, A.}
\newblock \bibinfo{journal}{\bibinfo{title}{The inefficiency of bitcoin}}.
\newblock {\emph{\JournalTitle{Economics Letters}}}
  \textbf{\bibinfo{volume}{148}}, \bibinfo{pages}{80--82},
  \url{https://doi.org/10.1016/j.econlet.2016.09.019} (\bibinfo{year}{2016}).

\bibitem{chakravarty_2020_the}
\bibinfo{author}{Chakravarty, M. M.~T.} \emph{et~al.}
\newblock \bibinfo{journal}{\bibinfo{title}{The extended utxo model}}.
\newblock {\emph{\JournalTitle{Financial Cryptography and Data Security}}}
  \bibinfo{pages}{525--539}, \url{https://doi.org/10.1007/978-3-030-54455-3_37}
  (\bibinfo{year}{2020}).

\bibitem{prezsol_2019_another}
\bibinfo{author}{Pérez-Solà, C.}, \bibinfo{author}{Delgado-Segura, S.},
  \bibinfo{author}{Navarro-Arribas, G.} \&
  \bibinfo{author}{Herrera-Joancomartí, J.}
\newblock \bibinfo{journal}{\bibinfo{title}{Another coin bites the dust: an
  analysis of dust in utxo-based cryptocurrencies}}.
\newblock {\emph{\JournalTitle{Royal Society Open Science}}}
  \textbf{\bibinfo{volume}{6}}, \bibinfo{pages}{180817},
  \url{https://doi.org/10.1098/rsos.180817} (\bibinfo{year}{2019}).

\bibitem{glenn_2005_cohort}
\bibinfo{author}{Glenn, N.~D.}
\newblock \emph{\bibinfo{title}{Cohort analysis}} (\bibinfo{publisher}{Sage
  Publications}, \bibinfo{year}{2005}).

\bibitem{mason_1973_some}
\bibinfo{author}{Mason, K.~O.}, \bibinfo{author}{Mason, W.~M.},
  \bibinfo{author}{Winsborough, H.~H.} \& \bibinfo{author}{Poole, W.~K.}
\newblock \bibinfo{journal}{\bibinfo{title}{Some methodological issues in
  cohort analysis of archival data}}.
\newblock {\emph{\JournalTitle{American Sociological Review}}}
  \textbf{\bibinfo{volume}{38}}, \bibinfo{pages}{242},
  \url{https://doi.org/10.2307/2094398} (\bibinfo{year}{1973}).

\bibitem{breslow_1983_multiplicative}
\bibinfo{author}{Breslow, N.~E.}, \bibinfo{author}{Lubin, J.~H.},
  \bibinfo{author}{Marek, P.} \& \bibinfo{author}{Langholz, B.}
\newblock \bibinfo{journal}{\bibinfo{title}{Multiplicative models and cohort
  analysis}}.
\newblock {\emph{\JournalTitle{Journal of the American Statistical
  Association}}} \textbf{\bibinfo{volume}{78}}, \bibinfo{pages}{1–12},
  \url{https://doi.org/10.2307/2287093} (\bibinfo{year}{1983}).

\bibitem{jiang_2016_cohort}
\bibinfo{author}{Jiang, D.} \emph{et~al.}
\newblock \bibinfo{journal}{\bibinfo{title}{Cohort query processing}}.
\newblock {\emph{\JournalTitle{Proceedings of the VLDB Endowment}}}
  \textbf{\bibinfo{volume}{10}}, \bibinfo{pages}{1--12},
  \url{https://doi.org/10.14778/3015270.3015271} (\bibinfo{year}{2016}).

\bibitem{omidvartehrani_2018_cohort}
\bibinfo{author}{Omidvar-Tehrani, B.}, \bibinfo{author}{Amer-Yahia, S.} \&
  \bibinfo{author}{Lakshmanan, L.~V.}
\newblock \bibinfo{journal}{\bibinfo{title}{Cohort representation and
  exploration}}.
\newblock {\emph{\JournalTitle{IEEE 5th International Conference on Data
  Science and Advanced Analytics (DSAA)}}}
  \url{https://doi.org/10.1109/dsaa.2018.00027} (\bibinfo{year}{2018}).

\bibitem{liu_2021_cryptocurrency}
\bibinfo{author}{Liu, Y.} \& \bibinfo{author}{Zhang, L.}
\newblock \bibinfo{journal}{\bibinfo{title}{Cryptocurrency valuation: An
  explainable ai approach}}.
\newblock {\emph{\JournalTitle{SSRN Electronic Journal}}}
  \url{10.2139/ssrn.3657986} (\bibinfo{year}{2021}).

\bibitem{day_2019_introducing}
\bibinfo{author}{Day, A.}, \bibinfo{author}{Medvedev, E.}, \bibinfo{author}{AK,
  N.} \& \bibinfo{author}{Price, W.}
\newblock \bibinfo{title}{Introducing six new cryptocurrencies in bigquery
  public datasets—and how to analyze them} (\bibinfo{year}{2019}).

\bibitem{a2021_introduction}
\bibinfo{title}{Introduction to partitioned tables  |  bigquery  |  google
  cloud}, \url{https://cloud.google.com/bigquery/docs/partitioned-tables}
  (\bibinfo{year}{2021}).

\bibitem{liu_2021_replication}
\bibinfo{author}{Liu, Y.}, \bibinfo{author}{Zhang, L.} \&
  \bibinfo{author}{Zhao, Y.}
\newblock \bibinfo{journal}{\bibinfo{title}{Replication data for: "deciphering
  bitcoin blockchain data by cohort analysis"}}.
\newblock {\emph{\JournalTitle{Harvard Dataverse}}}
  \url{https://doi.org/10.7910/DVN/XSZQWP} (\bibinfo{year}{2021}).

\bibitem{coinmetrics_2021_coinmetricsdata}
\bibinfo{author}{Coinmetrics}.
\newblock \bibinfo{title}{coinmetrics/data: Archives of data produced by
  coinmetrics.io}, \url{https://github.com/coinmetrics/data}
  (\bibinfo{year}{2021}).

\bibitem{shu_2020_realtime}
\bibinfo{author}{Shu, M.} \& \bibinfo{author}{Zhu, W.}
\newblock \bibinfo{journal}{\bibinfo{title}{Real-time prediction of bitcoin
  bubble crashes}}.
\newblock {\emph{\JournalTitle{Physica A: Statistical Mechanics and its
  Applications}}} \textbf{\bibinfo{volume}{548}}, \bibinfo{pages}{124477},
  \url{https://doi.org/10.1016/j.physa.2020.124477} (\bibinfo{year}{2020}).

\bibitem{li_2019_sentimentbased}
\bibinfo{author}{Li, T.~R.}, \bibinfo{author}{Chamrajnagar, A.~S.},
  \bibinfo{author}{Fong, X.~R.}, \bibinfo{author}{Rizik, N.~R.} \&
  \bibinfo{author}{Fu, F.}
\newblock \bibinfo{journal}{\bibinfo{title}{Sentiment-based prediction of
  alternative cryptocurrency price fluctuations using gradient boosting tree
  model}}.
\newblock {\emph{\JournalTitle{Frontiers in Physics}}}
  \textbf{\bibinfo{volume}{7}} (\bibinfo{year}{2019}).

\bibitem{giudici_2019_vector}
\bibinfo{author}{Giudici, P.} \& \bibinfo{author}{Pagnottoni, P.}
\newblock \bibinfo{journal}{\bibinfo{title}{Vector error correction models to
  measure connectedness of bitcoin exchange markets}}.
\newblock {\emph{\JournalTitle{Applied Stochastic Models in Business and
  Industry}}} \textbf{\bibinfo{volume}{36}}, \bibinfo{pages}{95--109},
  \url{https://doi.org/10.1002/asmb.2478} (\bibinfo{year}{2019}).

\bibitem{giudici_2021_libra}
\bibinfo{author}{Giudici, P.}, \bibinfo{author}{Leach, T.} \&
  \bibinfo{author}{Pagnottoni, P.}
\newblock \bibinfo{journal}{\bibinfo{title}{Libra or librae? basket based
  stablecoins to mitigate foreign exchange volatility spillovers}}.
\newblock {\emph{\JournalTitle{Finance Research Letters}}}
  \bibinfo{pages}{102054}, \url{https://doi.org/10.1016/j.frl.2021.102054}
  (\bibinfo{year}{2021}).

\bibitem{pagnottoni_2019_neural}
\bibinfo{author}{Pagnottoni, P.}
\newblock \bibinfo{journal}{\bibinfo{title}{Neural network models for bitcoin
  option pricing}}.
\newblock {\emph{\JournalTitle{Frontiers in Artificial Intelligence}}}
  \textbf{\bibinfo{volume}{2}}, \url{https://doi.org/10.3389/frai.2019.00005}
  (\bibinfo{year}{2019}).

\bibitem{karalevicius_2018_using}
\bibinfo{author}{Karalevicius, V.}, \bibinfo{author}{Degrande, N.} \&
  \bibinfo{author}{De~Weerdt, J.}
\newblock \bibinfo{journal}{\bibinfo{title}{Using sentiment analysis to predict
  interday bitcoin price movements}}.
\newblock {\emph{\JournalTitle{The Journal of Risk Finance}}}
  \textbf{\bibinfo{volume}{19}}, \bibinfo{pages}{56--75},
  \url{https://doi.org/10.1108/jrf-06-2017-0092} (\bibinfo{year}{2018}).

\bibitem{resta_2020_technical}
\bibinfo{author}{Resta, M.}, \bibinfo{author}{Pagnottoni, P.} \&
  \bibinfo{author}{De~Giuli, M.~E.}
\newblock \bibinfo{journal}{\bibinfo{title}{Technical analysis on the bitcoin
  market: Trading opportunities or investors’ pitfall?}}
\newblock {\emph{\JournalTitle{Risks}}} \textbf{\bibinfo{volume}{8}},
  \bibinfo{pages}{44}, \url{https://doi.org/10.3390/risks8020044}
  (\bibinfo{year}{2020}).

\bibitem{daniel_2020_the}
\bibinfo{author}{Daniel, K.}, \bibinfo{author}{Mota, L.},
  \bibinfo{author}{Rottke, S.} \& \bibinfo{author}{Santos, T.}
\newblock \bibinfo{journal}{\bibinfo{title}{The cross-section of risk and
  returns}}.
\newblock {\emph{\JournalTitle{The Review of Financial Studies}}}
  \textbf{\bibinfo{volume}{33}}, \bibinfo{pages}{1927--1979},
  \url{https://doi.org/10.1093/rfs/hhaa021} (\bibinfo{year}{2020}).

\bibitem{griffin_2020_is}
\bibinfo{author}{Griffin, J.~M.} \& \bibinfo{author}{Shams, A.}
\newblock \bibinfo{journal}{\bibinfo{title}{Is bitcoin really untethered?}}
\newblock {\emph{\JournalTitle{The Journal of Finance}}}
  \textbf{\bibinfo{volume}{75}}, \bibinfo{pages}{1913--1964},
  \url{https://doi.org/10.1111/jofi.12903} (\bibinfo{year}{2020}).

\bibitem{croman_2016_on}
\bibinfo{author}{Croman, K.} \emph{et~al.}
\newblock \bibinfo{journal}{\bibinfo{title}{On scaling decentralized
  blockchains}}.
\newblock {\emph{\JournalTitle{Financial Cryptography and Data Security}}}
  \bibinfo{pages}{106--125}, \url{https://doi.org/10.1007/978-3-662-53357-4_8}
  (\bibinfo{year}{2016}).

\bibitem{gervais_2016_on}
\bibinfo{author}{Gervais, A.} \emph{et~al.}
\newblock \bibinfo{journal}{\bibinfo{title}{On the security and performance of
  proof of work blockchains}}.
\newblock {\emph{\JournalTitle{Proceedings of the 2016 ACM SIGSAC Conference on
  Computer and Communications Security - CCS'16}}}
  \url{https://doi.org/10.1145/2976749.2978341} (\bibinfo{year}{2016}).

\bibitem{pagnotta_2021_decentralizing}
\bibinfo{author}{Pagnotta, E.~S.}
\newblock \bibinfo{journal}{\bibinfo{title}{Decentralizing money: Bitcoin
  prices and blockchain security}}.
\newblock {\emph{\JournalTitle{The Review of Financial Studies}}}
  \url{https://doi.org/10.1093/rfs/hhaa149} (\bibinfo{year}{2021}).

\bibitem{wang_2021_ess}
\bibinfo{author}{Wang, X.}, \bibinfo{author}{Wang, C.}, \bibinfo{author}{Zhou,
  K.} \& \bibinfo{author}{Cheng, H.}
\newblock \bibinfo{journal}{\bibinfo{title}{Ess: An efficient storage scheme
  for improving the scalability of bitcoin network}}.
\newblock {\emph{\JournalTitle{IEEE Transactions on Network and Service
  Management}}} \url{https://doi.org/10.1109/tnsm.2021.3127187}
  (\bibinfo{year}{2021}).

\bibitem{karau_2021_monetary}
\bibinfo{author}{Karau, S.}
\newblock \bibinfo{journal}{\bibinfo{title}{Monetary policy and
  cryptocurrencies}}.
\newblock {\emph{\JournalTitle{SSRN Electronic Journal}}}
  \url{https://doi.org/10.2139/ssrn.3949549} (\bibinfo{year}{2021}).

\bibitem{sciecon_2021_scieconutxo}
\bibinfo{author}{SciEcon}.
\newblock \bibinfo{title}{Sciecon/utxo: Deciphering bitcoin blockchain data by
  cohort analysis} (\bibinfo{year}{2021}).

\bibitem{luyaozhang_2021_scieconutxo}
\bibinfo{author}{Zhang, L.} \& \bibinfo{author}{Zhao, Y.}
\newblock \bibinfo{journal}{\bibinfo{title}{Sciecon/utxo: Preprint: Deciphering
  bitcoin blockchain data by cohort analysis}}.
\newblock {\emph{\JournalTitle{Zenodo}}}
  \url{https://doi.org/10.5281/zenodo.4708453} (\bibinfo{year}{2021}).

\end{thebibliography}

\section*{Acknowledgements} 

We have benefited from the comments by discussions at the SciEcon Research Accelerator Seminar. We thank two anonymous referees at Scientific Data for their insightful comments that helped us revise the manuscript. 

\section*{Author contributions statement}

Each author contributed equally to this research. Yulin Liu led the discussion on the blockchain mechanism and the economic meaning of UTXO data. Luyao Zhang designed the cohort analysis methods applying to UTXO-based blockchains. Yinhong Zhao queried and processed the data. Each author contributed significantly to the draft and revisions of the manuscript. 

\section*{Competing interests}

The authors have no competing interests to declare. 

\section*{Figures \& Tables}

\begin{figure}[ht]
\centering
\includegraphics[width=\linewidth]{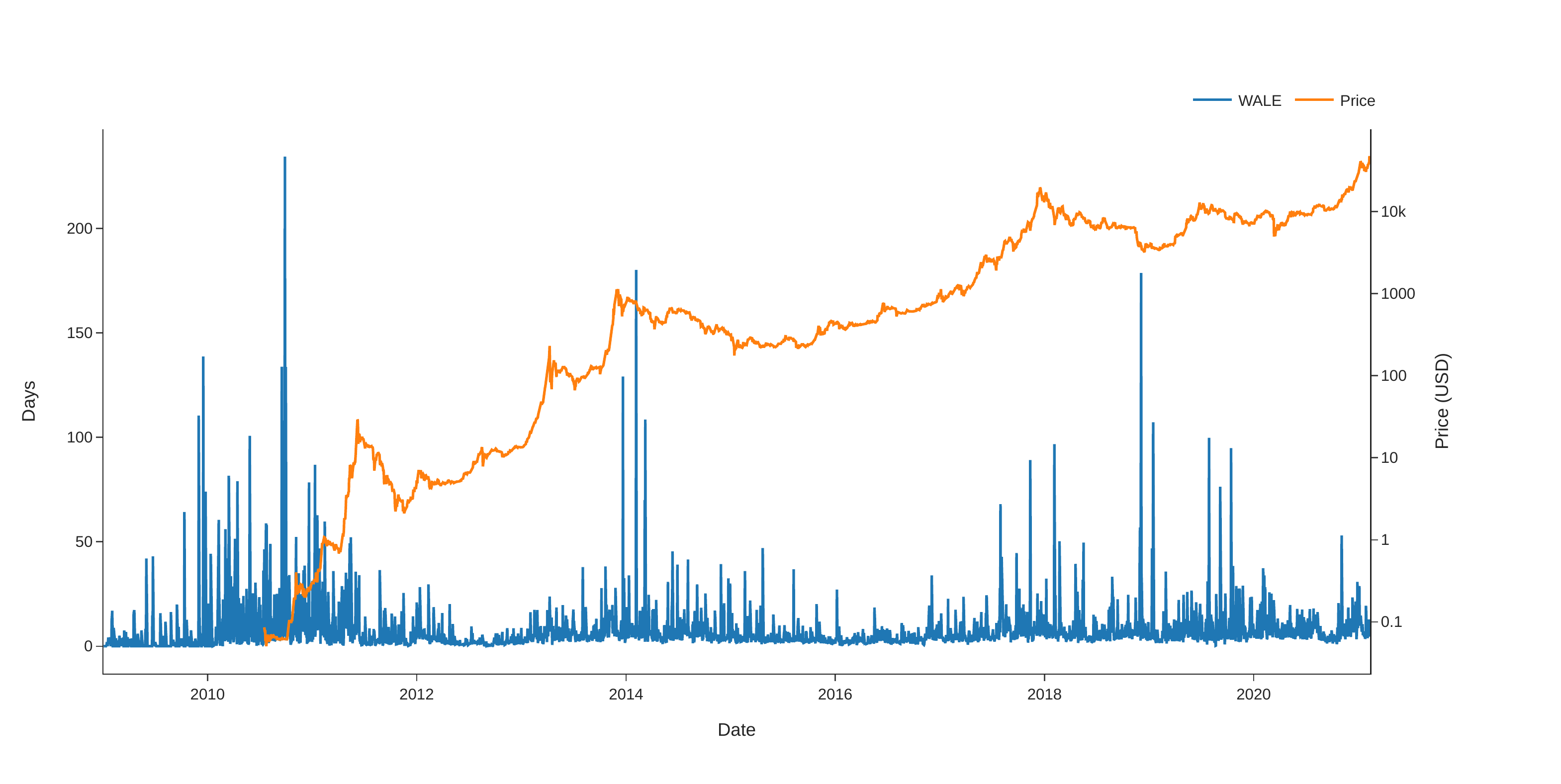}
\caption{Daily weighted average lifespan of Bitcoin STXOs and BTC price. The figure shows that the WAL of BTCs in STXOs attains a peak value when the BTC price is volatile. For example, the 2014 peak of WAL value closely followed the rocketing of BTC price from $\$100$ to $\$1000$ and its subsequent price collapse. This implies that older BTC become more active during market turmoil.}
\label{fig:4}
\end{figure}

\begin{table}[]
    \centering
    \begin{tabular}{cccccc}
    \hline \hline
        name & brief & start date & block time & link \\
        \hline
        Bitcoin & BTC & 2009-01-03 & $\approx$ 10 min & whitepaper~\cite{nakamoto_2008_bitcoin} \\
        Bitcoin Cash & BCH & 2017-07-31 & $\approx$ 10 min & \href{https://bitcoincash.org/}{link} \\
        Litecoin & LTC & 2011-10-23 & $\approx$ 2.5 min & \href{https://litecoin.org/}{link}\\
        Dogecoin & DOGE & 2013-12-06 & $\approx$ 1 min & \href{https://dogecoin.com/}{link} \\
        Dash & DASH & 2014-01-18 & $\approx$ 2.5 min & \href{https://www.dash.org/}{link} \\
        Zcash & ZEC & 2016-10-28 & $\approx$ 75 seconds & \href{https://z.cash/}{link} \\
        \hline \hline
    \end{tabular}
    \caption{Block time summary of UTXO-based cryptocurrencies. Data sources from white papers and websites of these cryptocurrencies.}
    \label{tab:blocktime}
\end{table}

\newpage
\section*{Appendix}
\subsection{Meta Data}
Table~\ref{table:A1} and Table~\ref{table:A2} present the meta data of the data we present on Bitcoin STXO and UTXO.
\begin{table}[h!]
\centering
\begin{tabular}{  p{1.5cm} p{14cm}} 
 \hline
 Name & Description\\ [0.5ex] 
 \hline
 date & Date on which cohort data were queried, in the format “\%Y/\%m/\%d”\\ 
 newborn & Number of UTXOs in BTC created on the given date \\
 dead & Number of UTXOs in BTC spent as inputs on the given date \\
 WAL & Weighted average lifespan of the UTXOs spent on the given date, defined as the average lifespan (the difference between the time when the output was spent and the time when the output were created) weighted by the number of UTXOs in BTC contained in the transaction outputs.\\
 -9 & Number of UTXOs in BTC spent on the given date that were created less than one day (< 1d) before\\
 -7 & Number of UTXOs in BTC spent on the given date that were created more than one day but less than one month (1d $\sim$ 1m) before\\ 
 -5 & Number of UTXOs in BTC spent on the given date that were created more than one month but less than three months  (1m $\sim$ 1q) before\\ 
 -3 & Number of UTXOs in BTC spent on the given date that were created more than three months but less than six months (1q $\sim$ 6m) before\\ 
 -1 & Number of UTXOs in BTC spent on the given date that were created more than six months but less than one year (6m $\sim$ 1y) before\\ 
 1 & Number of UTXOs in BTC spent on the given date that were created more than one year but less than two years (1y $\sim$ 2y) before\\ 
 3 & Number of UTXOs in BTC spent on the given date that were created more than two years but less than three years (2y $\sim$ 3y) before\\ 
 5 & Number of UTXOs in BTC spent on the given date that were created more than three years but less than four years (3y $\sim$ 4y) before\\
 7 & Number of UTXOs in BTC spent on the given date that were created more than four years but less than five years (4y $\sim$ 5y) before\\
 9 & Number of UTXOs in BTC spent on the given date that were created more than five years but less than ten years (5y $\sim$ 10y) before\\
 11 & Number of UTXOs in BTC spent on the given date that were created more than ten years (> 10y) before\\
 [1ex] 
 \hline
\end{tabular}
\caption{Meta Data for STXO Dataset}
\label{table:A1}
\end{table}

\begin{table}[h!]
\centering
\begin{tabular}{  p{1.5cm}  p{14cm} } 
 \hline
 Name & Description\\ [0.5ex] 
 \hline
 date & Date on which cohort data was queried, in the format “\%Y/\%m/\%d”\\ 
 -9 & Number of UTXOs in BTC still alive by the end of the given date that were created less than one day (< 1d) before\\
 -7 & Number of UTXOs in BTC still alive by the end of the given date that were created more than one day but less than one month (1d $\sim$ 1m) before\\ 
 -5 & Number of UTXOs in BTC still alive by the end of the given date that were created more than one month but less than three months  (1m $\sim$ 1q) before\\ -3 & Number of UTXOs still alive by the end of the given date that were created more than three months but less than six months (1q $\sim$ 6m) before\\ 
 -1 & Number of UTXOs in BTC still alive by the end of the given date that were created more than six months but less than one year (6m $\sim$ 1y) before\\
 1 & Number of UTXOs in BTC still alive by the end of the given date that were created more than one year but less than two years (1y $\sim$ 2y) before\\ 
 3 & Number of UTXOs in BTC still alive by the end of the given date that were created more than two years but less than three years (2y $\sim$ 3y) before\\ 
 5 & Number of UTXOs in BTC still alive by the end of the given date that were created more than three years but less than four years (3y $\sim$ 4y) before\\
 7 & Number of UTXOs in BTC still alive by the end of the given date that were created more than four years but less than five years (4y $\sim$ 5y) before\\
 9 & Number of UTXOs in BTC still alive by the end of the given date that were created more than five years but less than ten years (5y $\sim$ 10y) before\\
 11 & Number of UTXOs in BTC still alive by the end of the given date that were created more than ten years (> 10y) before\\
 [1ex] 
 \hline
\end{tabular}
\caption{Meta Data for UTXO Dataset}
\label{table:A2}
\end{table}
\par
Table~\ref{tab:datafiles} lists all the data files published in our GitHub depository, including the STXO and UTXO data for six cryptocurrencies including Bitcoin.

\begin{table}[]
    \centering
    \begin{tabular}{ccccc}
    \hline \hline
        name & brief & data end date & UTXO file name & STXO file name \\
        \hline
        Bitcoin & BTC  & 2021-02-10 & \href{https://github.com/SciEcon/UTXO/blob/main/bitcoin/BitcoinResultUTXO2021-02-10.csv}{BitcoinResultUTXO2021-02-10.csv} & \href{https://github.com/SciEcon/UTXO/blob/main/bitcoin/BitcoinResultSTXO2021-02-10.csv}{BitcoinResultSTXO2021-02-10.csv} \\
        Bitcoin Cash & BCH & 2020-12-31 & \href{https://github.com/SciEcon/UTXO/blob/main/bitcoin_cash/bitcoin_cashResultUTXO2020-12-31.csv}{bitcoin\_cashResultUTXO2020-12-31.csv} & 
        \href{https://github.com/SciEcon/UTXO/blob/main/bitcoin_cash/bitcoin_cashResultSTXO2020-12-31.csv}{bitcoin\_cashResultSTXO2020-12-31.csv} \\
        Litecoin & LTC & 2020-12-31 & \href{https://github.com/SciEcon/UTXO/blob/main/litecoin/litecoinResultUTXO2020-12-31.csv}{litecoinResultUTXO2020-12-31.csv} & 
        \href{https://github.com/SciEcon/UTXO/blob/main/litecoin/litecoinResultSTXO2020-12-31.csv}{litecoinResultSTXO2020-12-31.csv} \\
        Dogecoin & DOGE & 2020-12-31 & \href{https://github.com/SciEcon/UTXO/blob/main/dogecoin/dogecoinResultUTXO2020-12-31.csv}{dogecoinResultUTXO2020-12-31.csv} & 
        \href{https://github.com/SciEcon/UTXO/blob/main/dogecoin/dogecoinResultSTXO2020-12-31.csv}{dogecoinResultSTXO2020-12-31.csv} \\
        Dash & DASH & 2021-02-20 & \href{https://github.com/SciEcon/UTXO/blob/main/dash/dashResultUTXO2021-02-20.csv}{dashResultUTXO2021-02-20.csv} & 
        \href{https://github.com/SciEcon/UTXO/blob/main/dash/dashResultSTXO2021-02-20.csv}{dashResultSTXO2021-02-20.csv} \\
        Zcash & ZEC & 2020-12-31 & \href{https://github.com/SciEcon/UTXO/blob/main/zcash/zcashResultUTXO2020-12-31.csv}{zcashResultUTXO2021-02-20.csv} & 
        \href{https://github.com/SciEcon/UTXO/blob/main/zcash/zcashResultSTXO2020-12-31.csv}{zcashResultSTXO2021-02-20.csv} \\
        \hline \hline
    \end{tabular}
    \caption{File names of Altcoins UTXO and STXO data on GitHub}
    \label{tab:datafiles}
\end{table}

\subsection{Additional Visualizations}
\begin{figure}[ht]
\centering
\includegraphics[width=\linewidth]{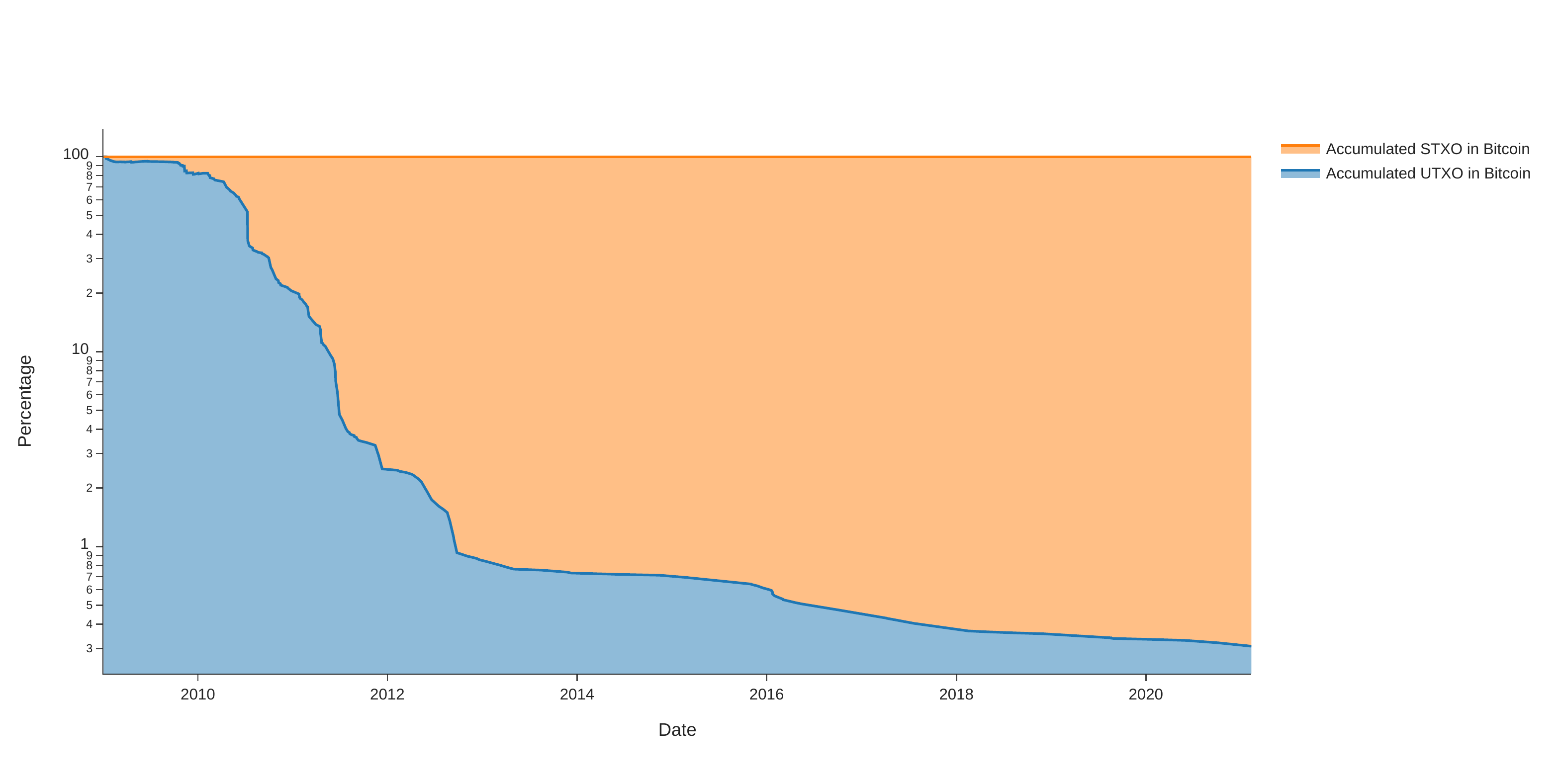}
\caption{Percentage of STXO and UTXO on Bitcoin}
\label{fig:A1}
\end{figure}

\begin{figure}[ht]
\centering
\includegraphics[width=\linewidth]{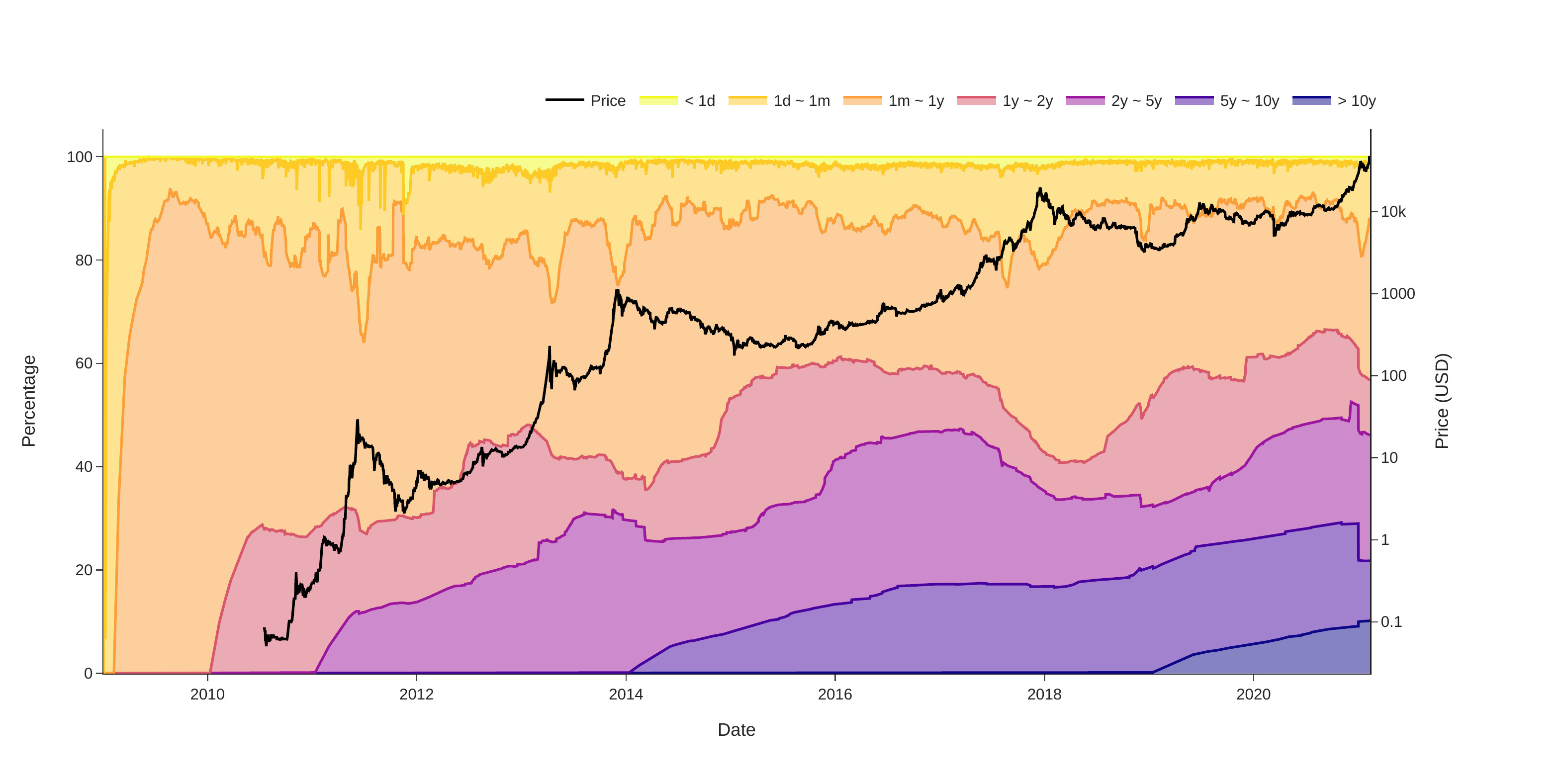}
\caption{Age distribution of UTXO on Bitcoin and the price of BTC}
\label{fig:A2}
\end{figure}

\begin{figure}[ht]
\centering
\includegraphics[width=\linewidth]{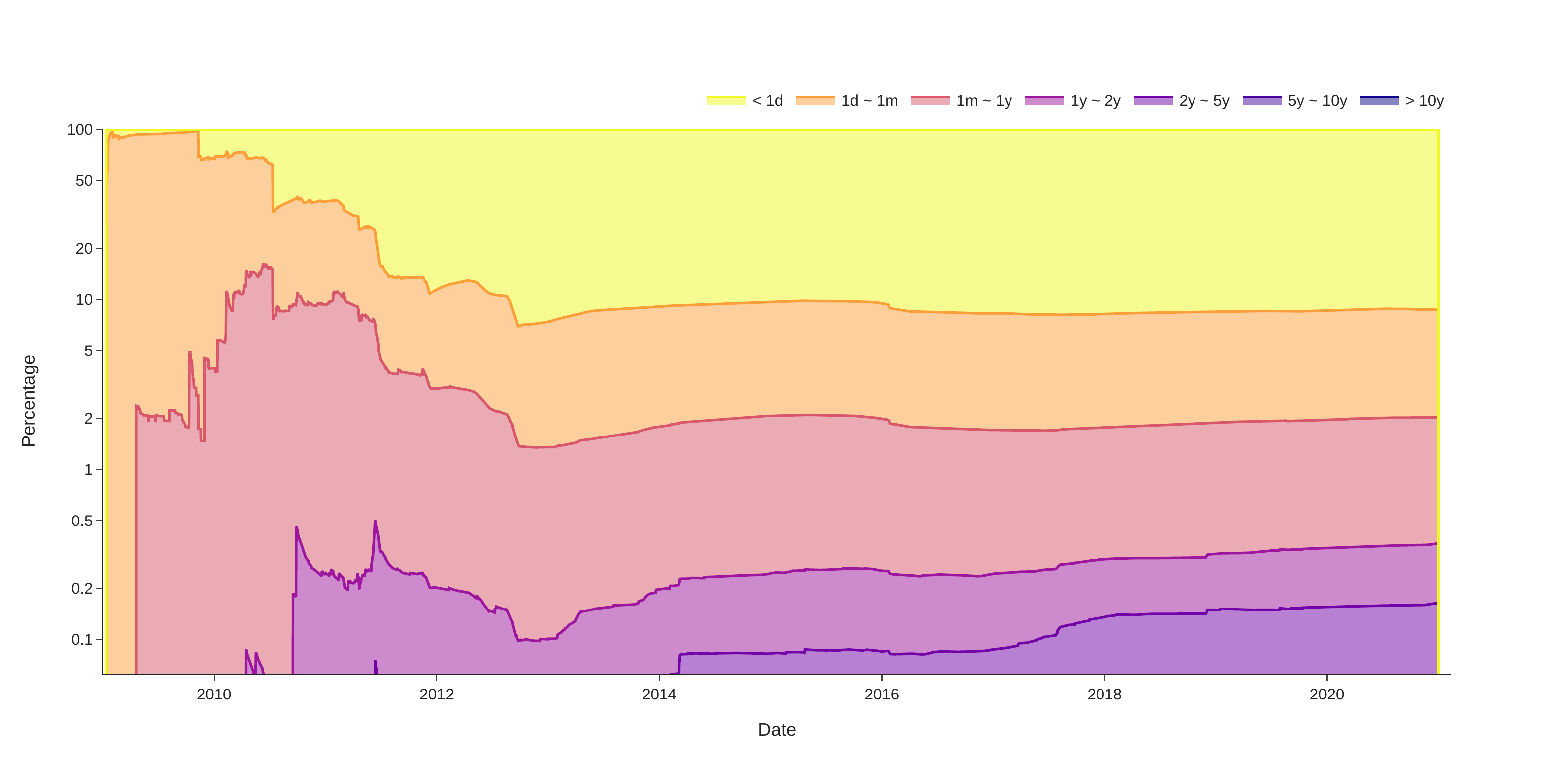}
\caption{Cumulative lifespan distribution of STXO on Bitcoin}
\label{fig:A3}
\end{figure}

\begin{figure}[ht]
\centering
\includegraphics[width=\linewidth]{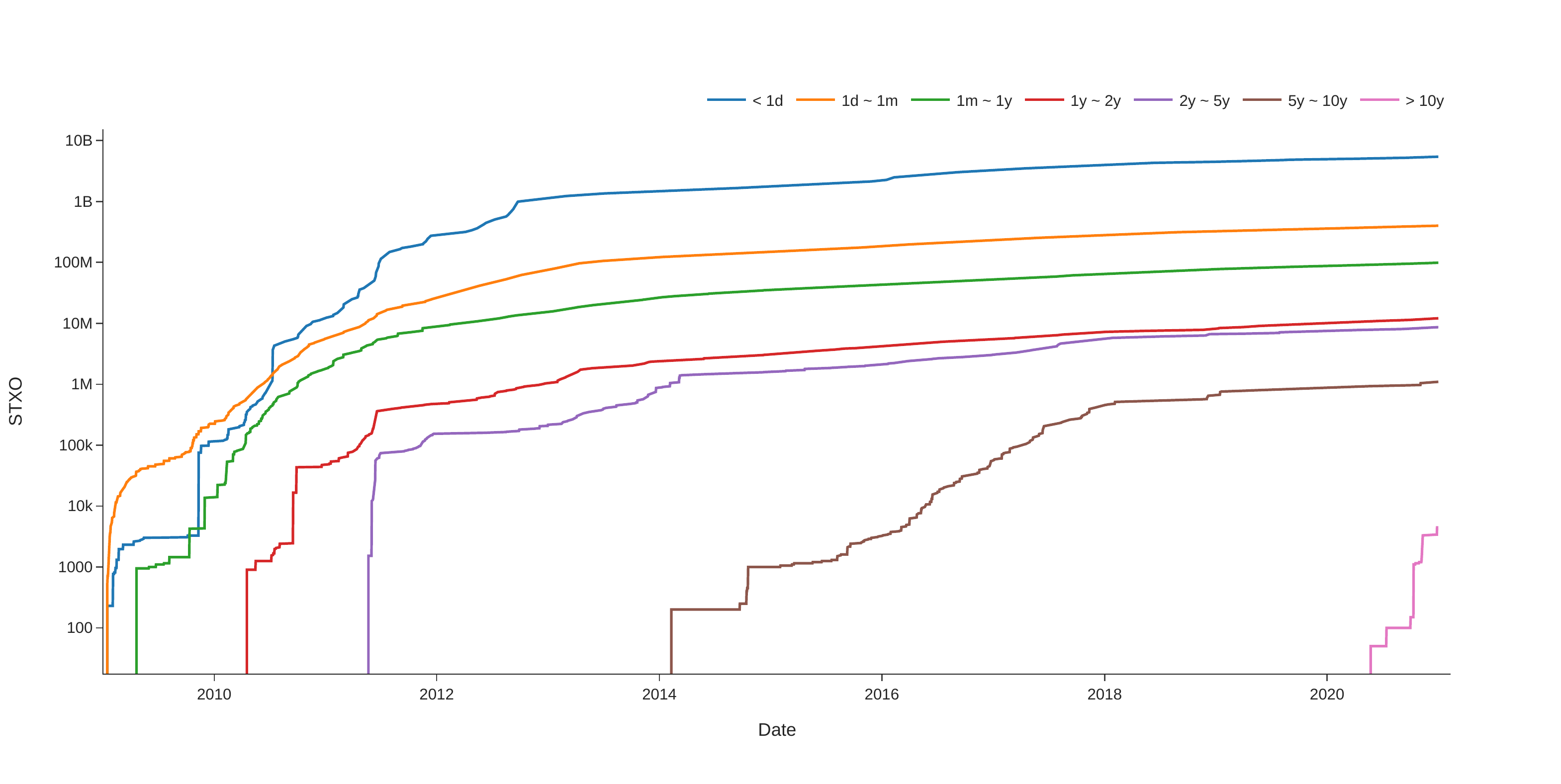}
\caption{Cumulative STXOs by lifespan on Bitcoin}
\label{fig:A4}
\end{figure}

\begin{figure}[ht]
\centering
\includegraphics[width=\linewidth]{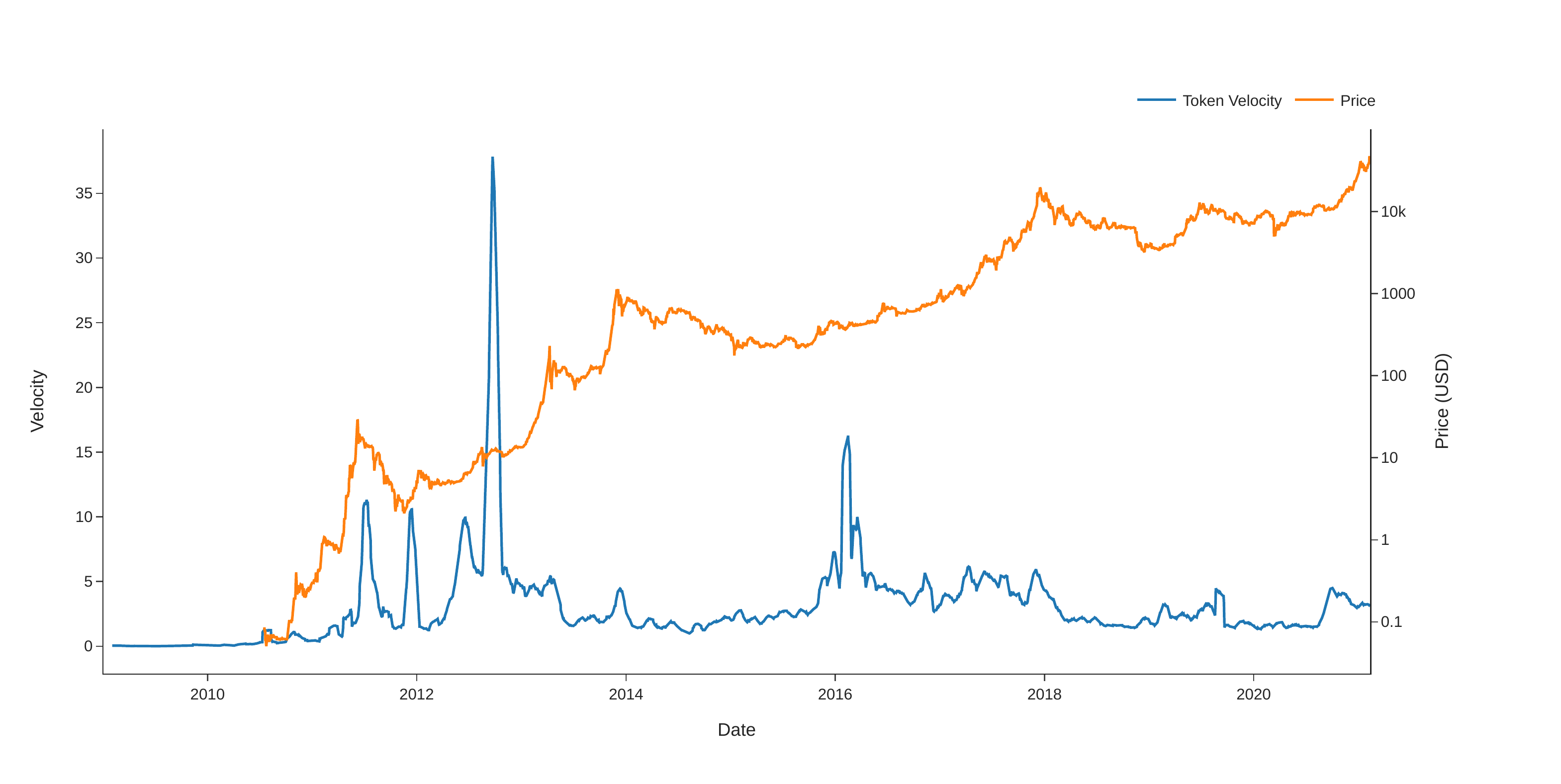}
\caption{Token velocity and price of BTC}
\label{fig:A5}
\end{figure}

Figures~\ref{fig:A1}, \ref{fig:A2}, \ref{fig:A3}, \ref{fig:A4}, and \ref{fig:A5} are several additional visualizations of the data we present here.

\end{document}